\documentclass[structabstract]{aa}  
\usepackage{graphicx}
\usepackage{txfonts}
\begin{document}
   \title{High-resolution images of five radio quasars at early cosmological epochs}

   \author{S.~Frey\inst{1,4}
           \and
           Z.~Paragi\inst{2,4}
	   \and
	   L.I.~Gurvits\inst{2,5}
           \and
	   D.~Cseh\inst{3}
	   \and
	   K.\'E.~Gab\'anyi\inst{1,4}
	   }

   \institute{F\"OMI Satellite Geodetic Observatory, P.O. Box 585, H-1592 Budapest, Hungary\\
              \email{frey@sgo.fomi.hu, gabanyik@sgo.fomi.hu}
         \and
              Joint Institute for VLBI in Europe, Postbus 2, 7990 AA Dwingeloo, The Netherlands\\
              \email{zparagi@jive.nl, lgurvits@jive.nl}
         \and
              Laboratoire Astrophysique des Interactions Multi-echelles (UMR 7158),
              CEA/DSM-CNRS-Universit\'e Paris Diderot, CEA Saclay, F-91191 Gif sur Yvette, France\\
              \email{david.cseh@cea.fr}
         \and
              MTA Research Group for Physical Geodesy and Geodynamics, P.O. Box 91, H-1521 Budapest, Hungary
         \and
              Institute of Space and Astronautical Science, Japan Aerospace Exploration Agency, 3-1-1 Yoshinodai Chuo-ku,
              Sagamihara, Kanagawa 252-5210, Japan}

   \date{Received Aug 9, 2010; accepted Sep 8, 2010}

 
  \abstract
   {Until now, there have only been seven quasars at $z>4.5$ whose the high-resolution radio structure had been studied in detail with Very Long Baseline Interferometry (VLBI) imaging.}
   {We almost double the number of VLBI-imaged quasars at these high redshifts with the aim of studying their redshift-dependent structural and physical properties in a larger sample.}
   {We observed five radio quasars (J0813+3508, J1146+4037, J1242+5422, J1611+0844, and J1659+2101) at $4.5 < z < 5$ with the European VLBI Network (EVN) at 1.6~GHz on 29 October 2008 and at 5~GHz on 22 October 2008. The angular resolution achieved ranges from 1.5 to 25 milli-arcseconds (mas), depending on the observing frequency, the position angle in the sky, and the source's celestial position.}
   {The sources are all somewhat extended on mas scales, but compact enough to be detected at both frequencies. With one exception of a flat-spectrum source (J1611+0844), their compact emission is characterised by a steep radio spectrum. We found no evidence of Doppler-boosted radio emission in the quasars in our sample. The radio structure of one of them (J0813+3508) is extended to $\sim$7$\arcsec$, which corresponds to 43~kpc projected linear size. Many of the highest redshift compact radio sources are likely to be young, evolving objects, far-away cousins of the powerful gigahertz peaked-spectrum (GPS) and compact steep-spectrum (CSS) sources that populate the Universe at lower redshifts.}  
   {}


   \maketitle
%

\section{Introduction}

Quasars at the highest redshifts place strong constraints on the early cosmological evolution of active galactic nuclei (AGNs) and the growth of their central supermassive ($\sim$$10^9$~$M_{\odot}$) black holes. The AGN activity observed at high $z$ indicates that feedback processes (e.g. Best et al. \cite{best05}) may have played an important role in the early galaxy and cluster evolution. The ultimate evidence for AGN jets is provided by Very Long Baseline Interferometry (VLBI) observations in the radio domain. 

Compact radio sources have flat spectra (with power-law spectral index $\alpha>-0.5$; $S\propto\nu^{\alpha}$, where $S$ is the flux density and $\nu$ the frequency) from the synchrotron self-absorption (Kellermann \& Pauliny-Toth \cite{kell69}). In the past, VLBI targets were traditionally selected by their flat overall radio spectrum to ensure detectability. Recent surveys that did not apply spectral selection criteria (e.g. Mosoni et al. \cite{moso06}; Frey et al. \cite{frey08a}), and VLBI observations of individual $z$$\sim$6 quasars (J0836+0054, Frey et al. \cite{frey03,frey05}; J1427+3312, Frey et al. \cite{frey08b}; Momjian et al. \cite{momj08}) indicate that there is a less-known steep-spectrum population of compact radio AGNs at high redshifts. These sources have so far been able to escape discovery for a variety of reasons: {\it (i)} their radio flux density is relatively low for VLBI, requiring high data rate phase-referenced observations; {\it (ii)} due to their steep spectra, they are generally not considered useful for VLBI experiments; {\it (iii)} for most of them, spectroscopic redshifts are simply not yet available. Indeed, the fine-scale radio structure of J0836+0054 ($z=5.77$) and J1427+3312 ($z=6.12$) could fortunately be revealed because their record-breaking redshifts made these sources attractive for high-resolution VLBI imaging.

The VLBI images of the highest-redshift, known radio-loud quasar J1427+3312 at 1.4~GHz (Momjian et al. \cite{momj08}) and 1.6~GHz (Frey et al. \cite{frey08b}) reveal a prominent double structure. The two resolved components separated by $\sim$28~mas ($\sim$160~pc) resemble a compact symmetric object (CSO, Wilkinson et al. \cite{wilk94}). (To calculate linear sizes and luminosities, we assume a flat cosmological model with $H_{\rm{0}}=70$~km~s$^{-1}$~Mpc$^{-1}$,  $\Omega_{\rm m}=0.3$, and $\Omega_{\Lambda}=0.7$ throughout this paper.) CSOs are a class of very young ($<10^4$~yr) sources typically found in radio galaxies at much lower redshifts ($z<1$). Apart from the structural similarity to CSOs, there are several indications of the youthfulness of this quasar: the steep radio spectrum coupled with the compact structure, the broad absorption lines, and the possible intrinsic X-ray absorption.

To our knowledge, there are only seven radio-loud AGNs at $z>4.5$ that have been imaged with VLBI prior to our experiment reported here (Table~\ref{z4.5-sources}). Four of them (J0906+6930, J1235$-$0003, J1430+4204, and J1451$-$1512) are compact, practically unresolved flat-spectrum radio sources, while others (J0836+0054, J0913+5919, and J1427+3312 with a double structure) have a compact or somewhat resolved appearance and a steep radio spectrum in the GHz frequency range (in the observer's frame). The double quasar J1205$-$0742 also listed in the table is a special case from our point of view, with measured brightness temperatures indicating extreme nuclear starbursts rather than radio-loud AGNs (Momjian et al. \cite{momj05}).


\begin{table}
  \caption[]{Quasars at $z>4.5$ imaged earlier with VLBI by increasing redshift.}
  \label{z4.5-sources}
\begin{tabular}{lccccc}        
\hline\hline                 
Source name   & $z$  & $\nu$ & Network & Peak     & Ref.  \\
              &      & GHz   &         & mJy/beam &       \\ 
\hline                       
J1235$-$0003  & 4.69 & 1.4   & VLBA    & 17       &  1 \\
J1205$-$0742N & 4.70 & 1.4   & VLBA    & 0.2      &  2 \\
J1205$-$0742S & 4.70 & 1.4   & VLBA    & 0.2      &  2 \\
J1430+4204    & 4.72 & 5     & EVN     & 161      &  3 \\
              &      & 15    & VLBA    & 159 \& 200      &  4 \\
J1451$-$1512  & 4.76 & 5     & EVN     & 50       &  5 \\
J0913+5919    & 5.11 & 1.4   & VLBA    & 19       &  1 \\
J0906+6930    & 5.47 & 15    & VLBA    & 115      &  6 \\
              &      & 43    & VLBA    & 42       &  6 \\
J0836+0054    & 5.77 & 1.6   & EVN     & 0.8      &  7 \\
              &      & 5     & EVN     & 0.3      &  8 \\
J1427+3312    & 6.12 & 1.4   & VLBA    & 1.0      &  9 \\
              &      & 1.6   & EVN     & 0.5      & 10 \\
              &      & 5     & EVN     & 0.2      & 10 \\
\hline   
\end{tabular}
\\
Notes: Col.~1 -- source name (J2000); Col.~2 -- spectroscopic redshift; Col.~3 -- observing frequency (GHz); 
Col.~4 -- interferometer array (EVN: European VLBI Network, VLBA: Very Long Baseline Array);
Col.~5 -- peak brightness (mJy/beam); 
Col.~6 -- references (1: Momjian et al. \cite{momj04}; 2: Momjian et al. \cite{momj05}; 3: Paragi et al. \cite{para99}; 4: Veres et al. \cite{vere10}; 5: L.I.~Gurvits et al., in preparation; 6: Romani et al. \cite{roma04}, 7: Frey et al. \cite{frey03}; 8: Frey et al. \cite{frey05}; 9: Momjian et al. \cite{momj08}; 10: Frey et al. \cite{frey08b}).
\end{table}

Our goal was to substantially increase the number of radio-loud AGNs at $z>4.5$ imaged with VLBI. One could expect to find either more ``classical'' core--jet sources or other steep-spectrum quasars, possibly with CSO-like double structures. If the latter are found, in a decade-long term, VLBI monitoring would eventually allow measurements of the component expansions and thus facilitate direct estimations of the kinematic age of the sources. Otherwise, the compact core--jet sources are potentially valuable additions for comparing of the mas-scale structures of quasars at low and high redshift. Efforts to use classical cosmological tests -- the angular size--redshift relation (e.g. Gurvits et al. \cite{gurv99}) and the apparent proper motion--redshift relation (e.g. Kellermann et al. \cite{kell99}) -- would also benefit from data on an increased sample of extremely distant quasars, since the predictions of the various cosmological world models are markedly different at the highest redshifts. Despite the practical difficulties, there is continuous interest in these tests in the community (e.g. Sahni \& Starobinsky \cite{sahn00}; Vishwakarma \cite{vish00,vish01}; Lima \& Alcaniz \cite{lima02}; Chen \& Ratra \cite{chen03}; Jackson \cite{jack04,jack08}). 

In this paper, we report on our dual-frequency VLBI imaging observations of five previously unexplored radio-loud quasars at $4.5 < z < 5$. Our sample selection method is described in Sect.~\ref{sample}. The experiments and the data reduction are explained in Sect.~\ref{experiment}. The observed radio properties of the sources are given in Sect.~\ref{results}. Our results are discussed in Sect.~\ref{discussion}. Conclusions are drawn in Sect.~\ref{conclusion}.

\section{Target selection}
\label{sample}

For the high-resolution VLBI observations, we choose five quasars from the Sloan Digital Sky Survey (SDSS) Data Release 5 (DR5) quasar catalogue (Schneider et al. \cite{schn07}). All of them are identified with an unresolved ($<5\arcsec$) radio source in the Very Large Array (VLA) Faint Images of the Radio Sky at Twenty-centimeter (FIRST) survey\footnote{\tt{http://sundog.stsci.edu}} (White et al. \cite{whit97}), with 1.4-GHz total flux densities 8.8~mJy $\le S_{\rm 1.4} \le$ 28.8~mJy. These five quasars, together with J0913+5919 (already studied with VLBI by Momjian et al. \cite{momj04}; see Table~\ref{z4.5-sources}), are the only such radio quasars in the northern hemisphere in the Schneider et al. (\cite{schn07}) catalogue with $z>4.5$ and $S_{\rm 1.4}>5$~mJy. Important is that the (otherwise unknown) radio spectral index of the sources was not used as a selection criterion. 

\begin{figure}
\centering
  \includegraphics[bb=45 165 570 640,width=80mm,angle=0,clip= ]{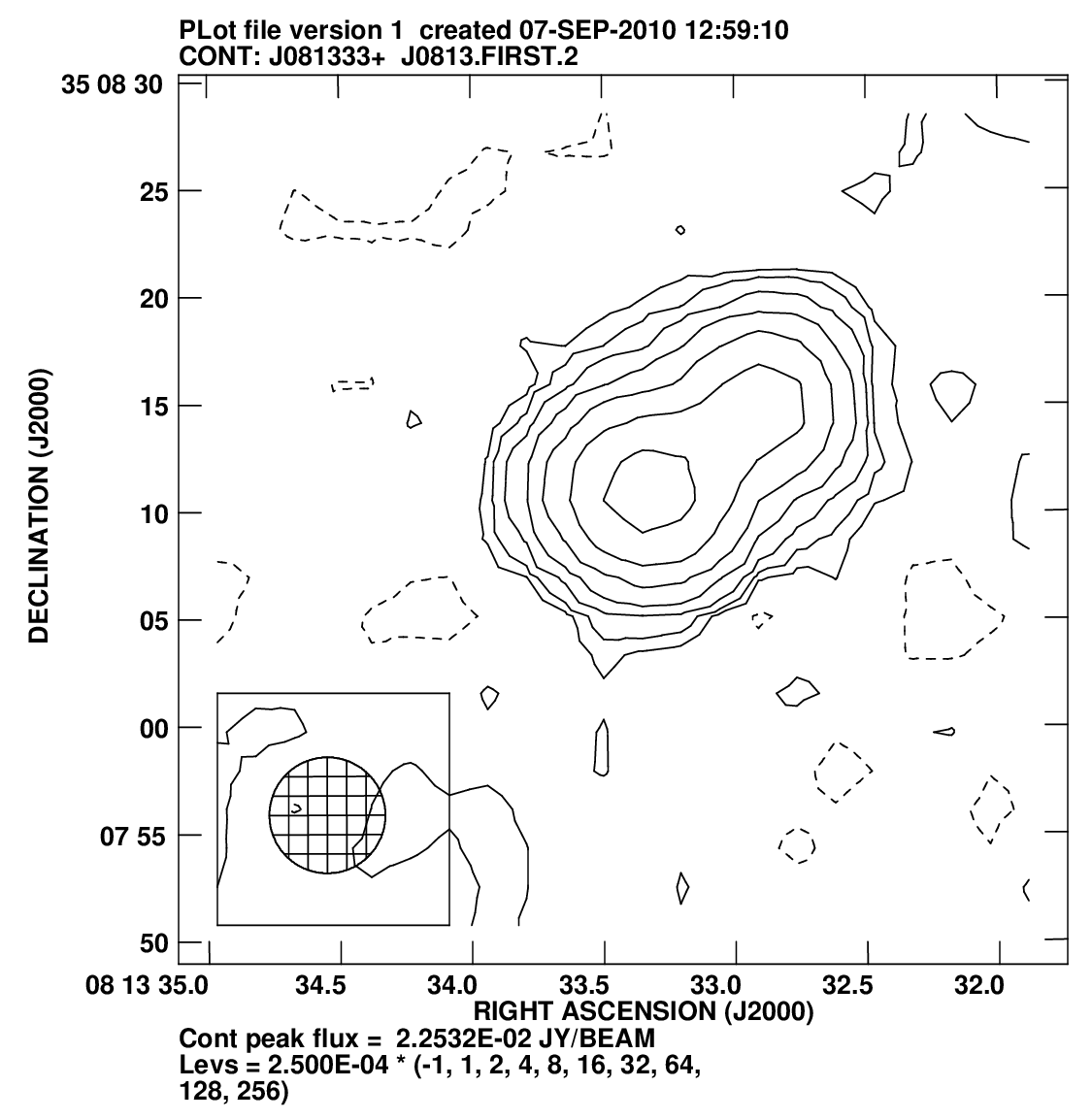}
  \caption{
The 1.4-GHz image of J0813+3508 from the VLA FIRST survey (White et al. \cite{whit97}). The peak brightness is 22.5~mJy/beam, the lowest contour levels are drawn at $\pm0.25$~mJy/beam, the positive contour levels increase by a factor of 2. The circular Gaussian restoring beam size is $5\farcs4$ (FWHM), as indicated in the lower-left corner. The northwestern radio companion (top right) is catalogued as another source in FIRST, $6\farcs8$ away from the quasar (in the centre). 
   }
  \label{first}
\end{figure}

One of the sources (J0813+3508, $z=4.92$) has a close ($<7\arcsec$), optically unidentified radio companion in the FIRST catalogue, which appears compact on an arcsecond scale in the 1.4-GHz VLA FIRST image (Fig.~\ref{first}). We also included this object as the sixth target in our VLBI experiment, in the hope that any possible relation between the two apparently nearby sources could be studied. The basic parameters of our target sources are listed in Table~\ref{targets}.

Our recent experience with a larger sample of SDSS/FIRST quasars (Mosoni et al. \cite{moso06}; Frey et al. \cite{frey08a}) indicates that the sources identified as optical quasars {\em and} unresolved FIRST objects with $S_{\rm 1.4}>20$~mJy have a nearly 90\% chance to be successfully detected with the European VLBI Network (EVN) at 5~GHz, using the SDSS coordinates as a priori values. Although most of the sources are weaker in the present sample, we used longer integration times in the VLBI experiment to ensure safe detections.


\begin{table}
  \caption[]{Our VLBI targets, five SDSS/FIRST radio quasars at $z>4.5$, and a close radio companion of one of them.}
  \label{targets}
\begin{tabular}{lcccc}        
\hline\hline                 
Source coordinates        & $z$  & $r$  & FIRST peak & $S_{\rm 1.4}$  \\
                          &      &      & mJy/beam   & mJy            \\ 
\hline                       
08 13 33.32  +35 08 10.8  & 4.92 & 20.8 &  23.2      & 25.2            \\
08 13 32.89  +35 08 14.9$^{*}$ & -    & -    &  11.8      & 11.8            \\
11 46 57.79  +40 37 08.6  & 5.01 & 21.0 &  12.5      & 12.5            \\
12 42 30.58  +54 22 57.3  & 4.73 & 20.9 &  19.7      & 20.2            \\
16 11 05.64  +08 44 35.4  & 4.54 & 19.7 &   8.8      &  8.8            \\
16 59 13.23  +21 01 15.8  & 4.78 & 21.5 &  28.7      & 28.8            \\
\hline   
\end{tabular}
\\
Notes: Col.~1 -- a priori source J2000 right ascension ($^{\rm h}$ $^{\rm m}$ $^{\rm s}$) and declination ($\degr$ $\arcmin$ $\arcsec$) from SDSS; Col.~2 -- spectroscopic redshift; Col.~3 -- SDSS $r$ magnitude (Schneider et al. \cite{schn07}); 
Col.~4 -- FIRST peak brightness (mJy/beam); Col.~5 -- FIRST integral 1.4-GHz flux density (mJy).\\
$^{*}$ optically unidentified radio companion of the previous source; coordinates from FIRST
\end{table}

\section{VLBI observations and data reduction}
\label{experiment}

We observed our targets with the EVN at 1.6~GHz on 29 October 2008 and at 5~GHz on 22 October 2008. Due to the high redshifts, the observed frequencies correspond to $\sim$9$-$10~GHz and $\sim$28$-$30~GHz in the rest frame of the quasars. At a recording rate of 1024~Mbit~s$^{-1}$, ten antennas of the EVN participated in the experiment at 1.6 GHz: Effelsberg (Germany), Jodrell Bank Mk2 (UK), Medicina, Noto (Italy), Toru\'n (Poland), Onsala (Sweden), Sheshan, Nanshan (P.R. China), Robledo (Spain), and the phased array of the 14-element Westerbork Synthesis Radio Telescope (WSRT, The Netherlands). The eight participating antennas in the 5-GHz experiment were Effelsberg, Jodrell Bank Mk2, Medicina, Noto, Toru\'n, Sheshan, Nanshan, and the WSRT. On both days, the observations lasted for a total of 11 h. Eight intermediate frequency channels (IFs) were used in both left and right circular polarisations. The total bandwidth was 128~MHz per polarisation. The correlation of the recorded VLBI data took place at the EVN Data Processor at the Joint Institute for VLBI in Europe (JIVE), Dwingeloo, the Netherlands.

All the target sources were observed in phase-reference mode. This allows us to increase the coherent integration time spent on the source and thus to improve the sensitivity of the observations. Phase-referencing involves regularly interleaving observations between the target source and a nearby bright and compact reference source (e.g. Beasley \& Conway \cite{Beas95}). The delay, delay rate, and phase solutions derived for the phase-reference calibrator were interpolated and applied for the respective target within the cycle time of $\sim$7 minutes. The target sources were observed for $\sim$4.5-minute intervals in each cycle. The exception was the pair of sources, J0813+3508 and its optically unidentified apparent companion (Table~\ref{targets}; Fig.~\ref{first}). They were observed using the same phase-reference source, for $\sim$2.5 minutes subsequently in each cycle. To achieve nearly the same total on-source integration time as for the other targets ($\sim$60~min), more cycles were scheduled for this pair.

The suitable phase-reference calibrator sources (J0815+3635, J1146+3958, J1253+5301, J1608+1029, and J1656+1826) were selected from the Very Long Baseline Array (VLBA) Calibrator Survey\footnote{{\tt http://www.vlba.nrao.edu/astro/calib/index.shtml}}. The angular separations between the calibrators and the corresponding targets range from $0\fdg64$ to $2\fdg65$ (Table \ref{image-param}). The positional uncertainties of the calibrators in the International Celestial Reference Frame (ICRF) are $0.4-4.3$~mas.

The US National Radio Astronomy Observatory (NRAO) Astronomical Image Processing System (AIPS; Diamond \cite{Diam95}) was used for the data calibration. The visibility amplitudes were calibrated using the antenna gains, and the system temperatures regularly measured at the antennas during the experiments. Fringe-fitting was performed for the five calibrators mentioned above, and the fringe-finder sources (J0555+3948, J0927+3902, J1159+2914, and J1331+3030) using 3-min solution intervals. The data were exported to the Caltech Difmap package (Shepherd et al. \cite{shep94}) for imaging. The conventional mapping procedure involving several iterations of CLEANing and phase (then amplitude) self-calibration resulted in the images and brightness distribution models for the calibrators. Overall antenna gain correction factors ($\sim$10\% or less) were determined and applied to the visibility amplitudes in AIPS. Then fringe-fitting was repeated in AIPS, now taking the clean component models of the phase-reference calibrator sources  into account. The residual phase corrections resulted from their non-pointlike structure were considered this way. The solutions obtained were interpolated and applied to the target source data. The visibility data of the target sources, unaveraged in time and frequency, were also exported to Difmap for imaging. The naturally weighted images at 1.6~GHz and 5~GHz (Fig.~\ref{images}) were made after several cycles of CLEANing in Difmap. Phase-only self-calibration was applied for the brighter sources (i.e. when the sum of the CLEAN component flux densities exceeded $\sim$$10$~mJy) over time intervals not shorter than the length of scans spent on the sources (see Table~\ref{image-param}). In the cases where phase self-calibration was not attempted at all, we expect a loss of coherence of about 5\% (cf. Mart\'{\i}-Vidal et al. \cite{mart10}) in the phase-referencing process, which may lead to an underestimate of the flux density values by this factor. In the images, the lowest contours are drawn at $\sim$3$\sigma$ image noise levels. The expected theoretical thermal noise values were $15-22$~$\mu$Jy/beam ($1\sigma$), assuming no data loss during the experiment. The image parameters are summarised in Table~\ref{image-param}.

\begin{table*}
  \caption[]{VLBI image parameters for Fig.\ref{images}.}
  \label{image-param}
\centering 
\begin{tabular}{ccccccc}        
\hline\hline                 
Source name & $\phi$  & $\nu$ & Peak & Contour & \multicolumn{2}{c}{Restoring beam}    \\
            & $\degr$ & GHz     & \multicolumn{2}{c}{mJy/beam}   & mas$\times$mas  & PA ($\degr$) \\ 
\hline                       
J0813+3508  & 1.50    & 1.6       & 11.4           & 0.20           &   $19.0 \times 4.9$     &  5.3 \\
            &         & 5         &  4.0$^{*}$     & 0.15           &   $ 5.4 \times 1.5$     &  8.6 \\
J1146+4037  & 0.64    & 1.6       & 14.7           & 0.30           &   $20.1 \times 4.4$     &  5.4 \\
            &         & 5         &  6.0$^{*}$     & 0.20           &   $ 5.1 \times 1.5$     &  3.6 \\
J1242+5422  & 2.09    & 1.6       & 12.1           & 0.60           &   $17.1 \times 4.5$     & 22.2 \\
            &         & 5         &  7.9           & 0.15           &   $ 4.5 \times 1.5$     & 22.7 \\
J1611+0844  & 1.83    & 1.6       & 11.1           & 1.00           &   $25.7 \times 3.4$     &  9.6 \\
            &         & 5         &  9.4           & 0.25           &   $ 7.1 \times 1.5$     &  9.1 \\
J1659+2101  & 2.65    & 1.6       & 16.6           & 0.30           &   $22.4 \times 3.7$     & 11.6 \\
            &         & 5         &  4.2           & 0.50           &   $ 5.8 \times 1.4$     & 11.7 \\
\hline   
\end{tabular}
\\
Notes: Col.~1 -- source name (J2000); 
Col.~2. -- angular separation from the phase-reference calibrator source ($\degr$);
Col.~3 -- observing frequency (GHz); Col.~4 -- peak
brightness (mJy/beam); cases where phase self-calibration was not applied are marked with asterisks; Col.~5 -- lowest contour level (mJy/beam) corresponding
to $\sim$3$\sigma$ image noise; Col.~6 -- Gaussian restoring beam size (mas$\times$mas);
Col.~7 -- restoring beam major axis position angle ($\degr$) measured from north through east.
\end{table*}

\begin{figure*}[!t]
\centering
  \includegraphics[bb=68 170 520 626,width=75mm,angle=270,clip= ]{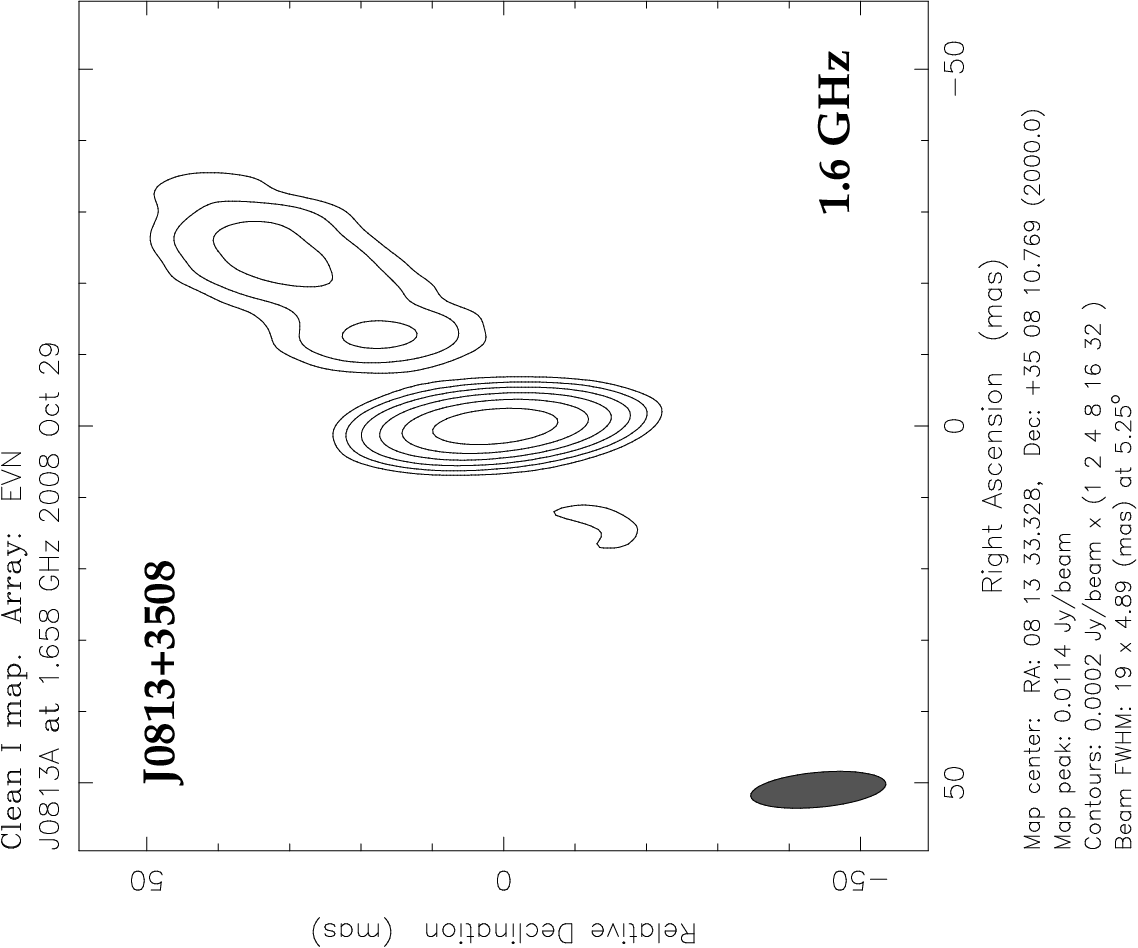}
  \includegraphics[bb=68 170 520 626,width=75mm,angle=270,clip= ]{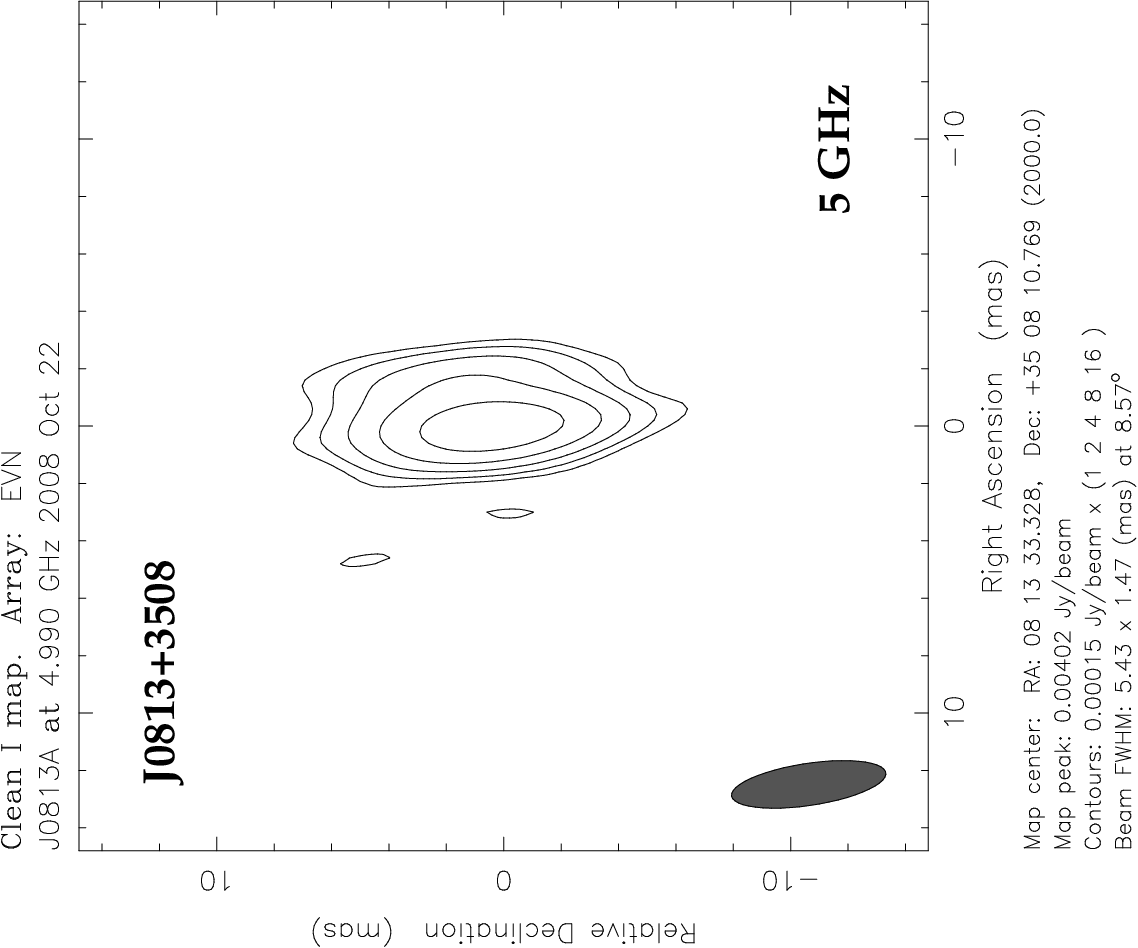}
  \includegraphics[bb=68 170 520 626,width=75mm,angle=270,clip= ]{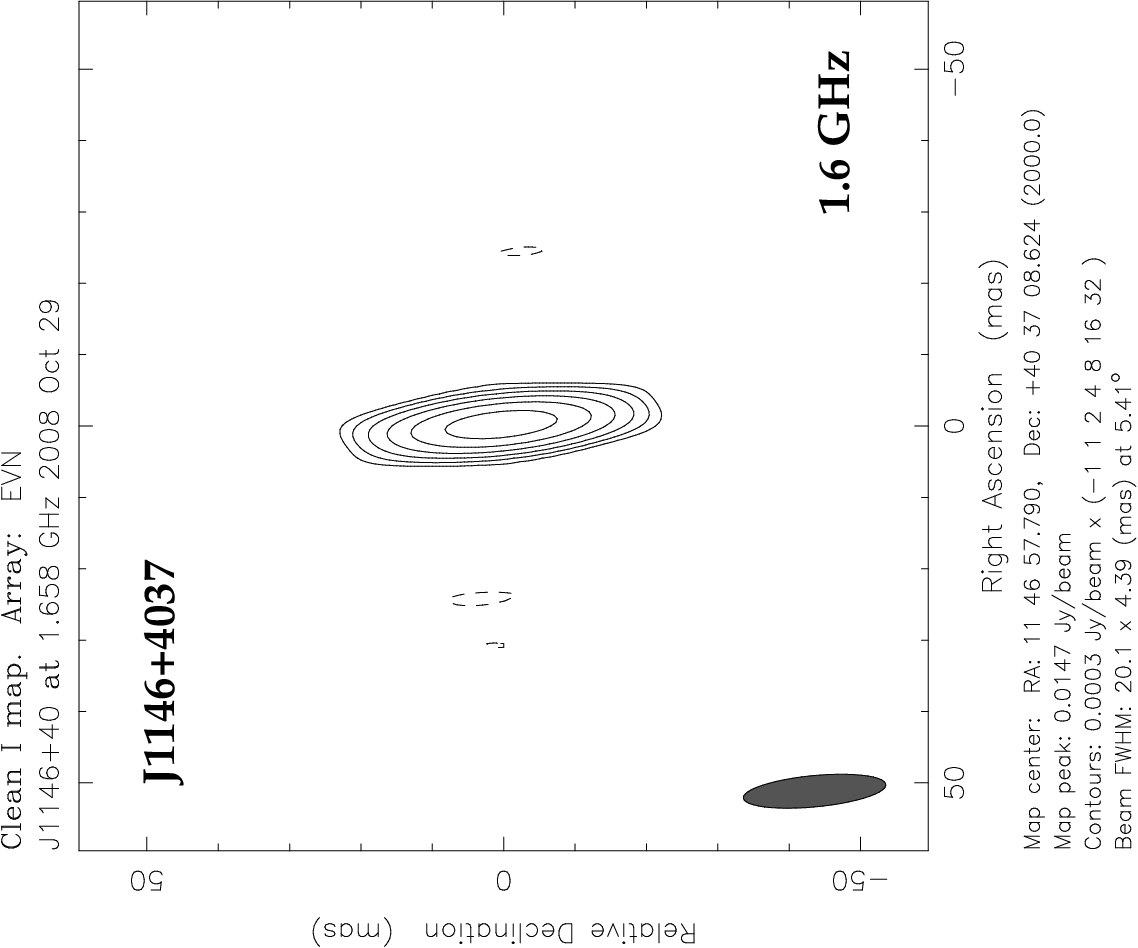}
  \includegraphics[bb=68 170 520 626,width=75mm,angle=270,clip= ]{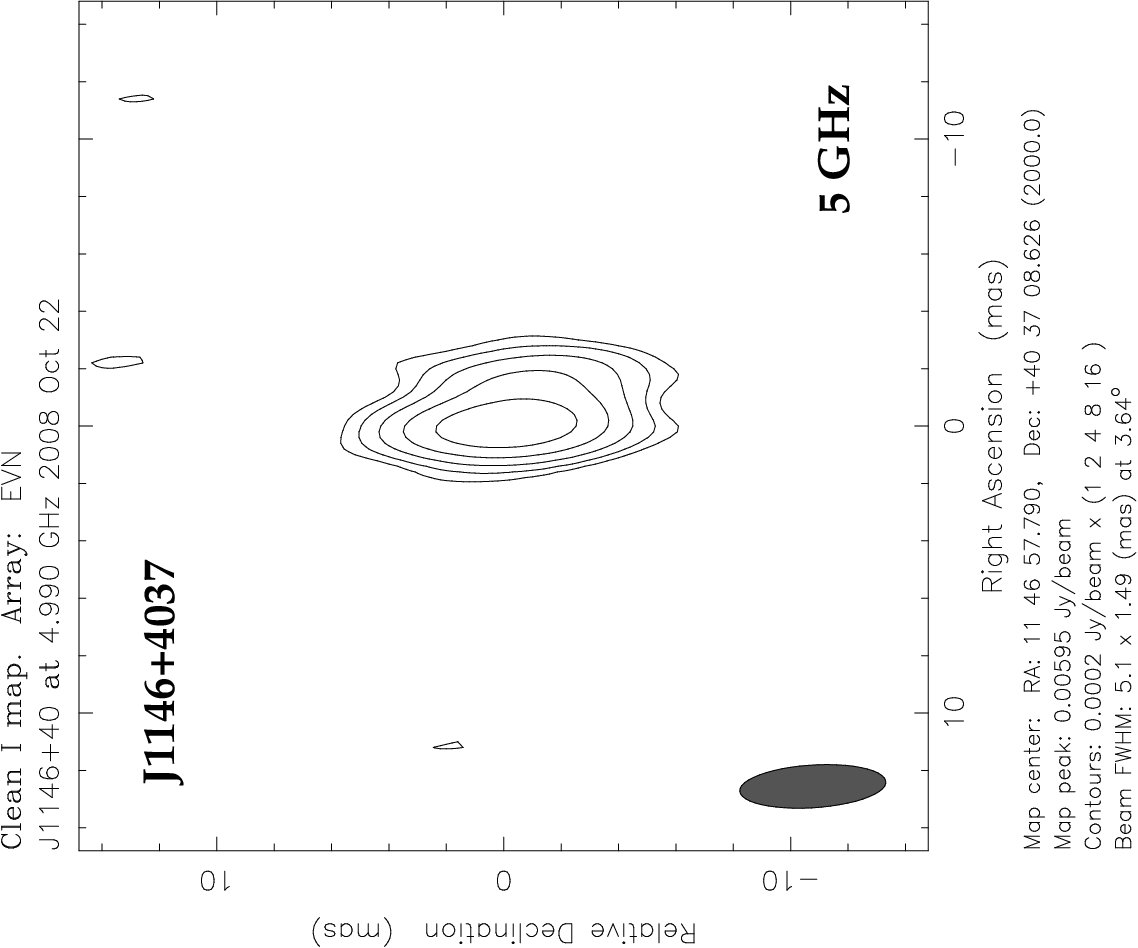}
  \includegraphics[bb=68 170 520 626,width=75mm,angle=270,clip= ]{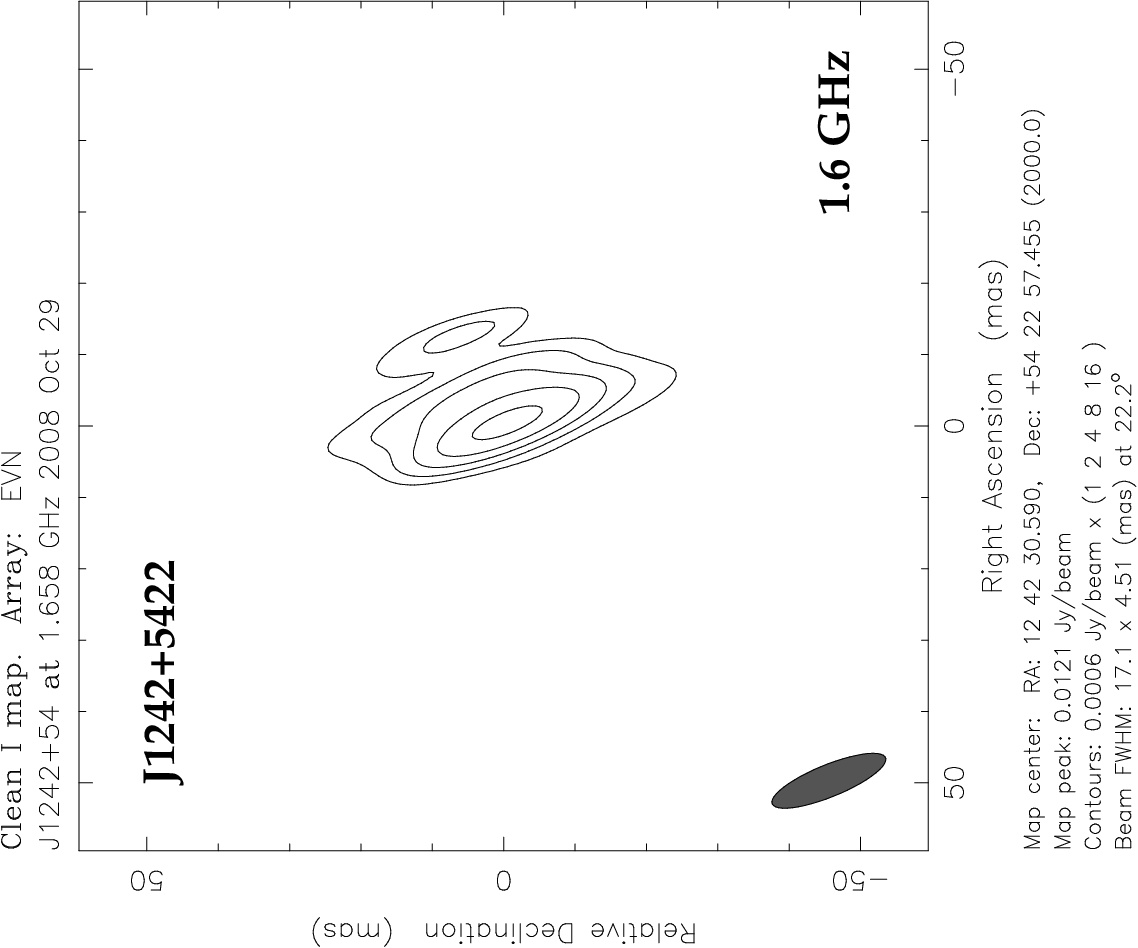}
  \includegraphics[bb=68 170 520 626,width=75mm,angle=270,clip= ]{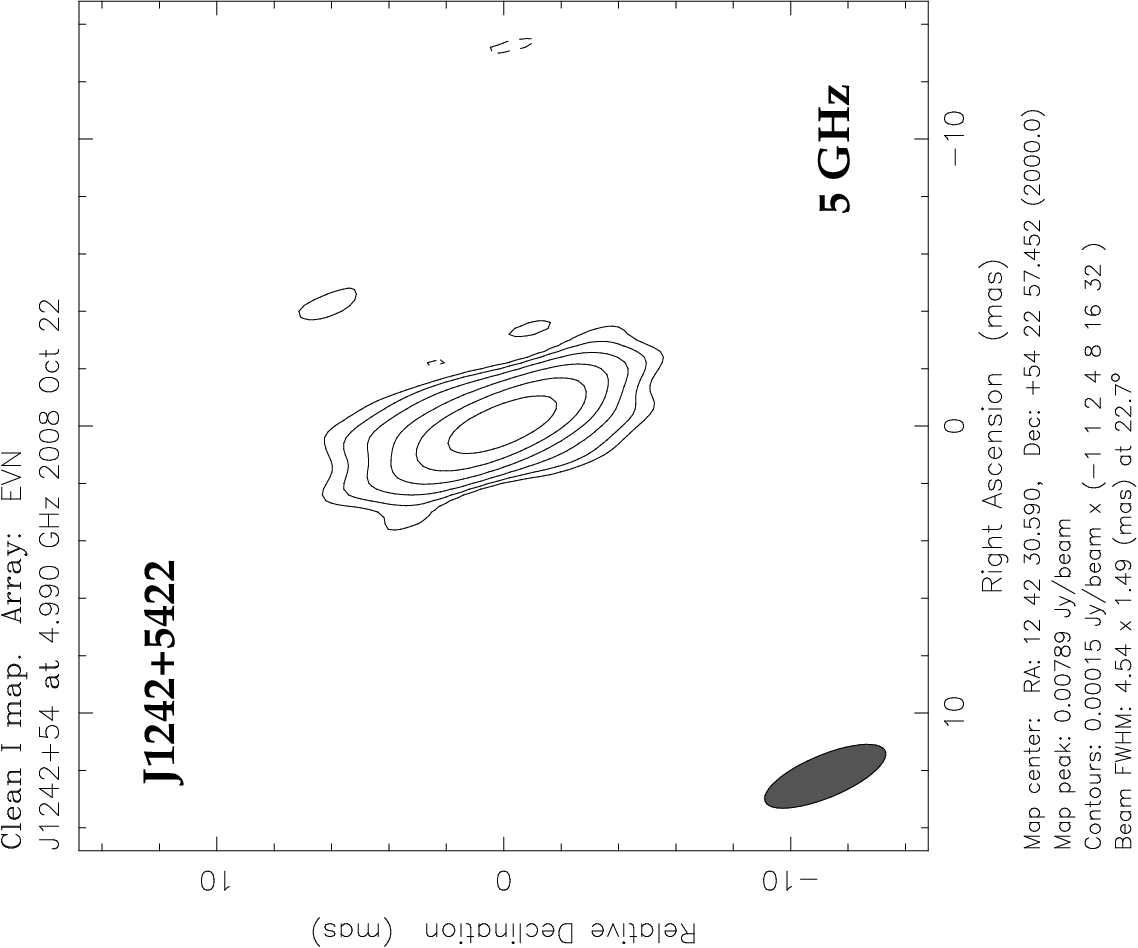}
  \caption{
The naturally weighted 1.6-GHz {\it (left column)} and 5-GHz {\it (right column)} VLBI images of the quasars. The image parameters (peak brightness, lowest contour level corresponding to $\sim$3$\sigma$ image noise, restoring beam size, and orientation) are listed in Table~\ref{image-param}. The full width at half maximum (FWHM) of the Gaussian restoring beam is indicated with an ellipse in the lower-left corners. The positive contour levels increase by a factor of 2. The coordinates are related to the brightness peak of which the absolute position is given in Table~\ref{coords}.
   }
  \label{images}
\end{figure*}

\addtocounter{figure}{-1}
\begin{figure*}
\centering
  \includegraphics[bb=68 170 520 626,width=75mm,angle=270,clip= ]{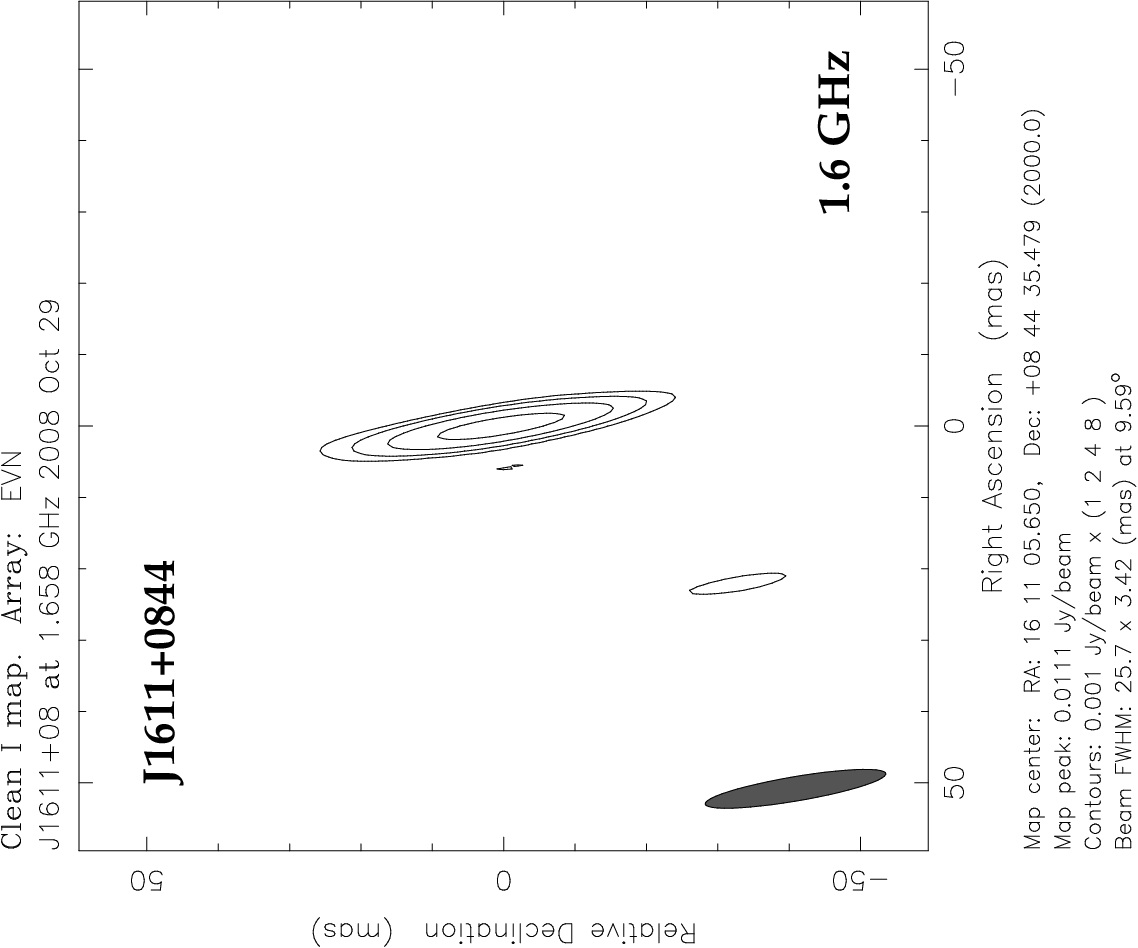}
  \includegraphics[bb=68 170 520 626,width=75mm,angle=270,clip= ]{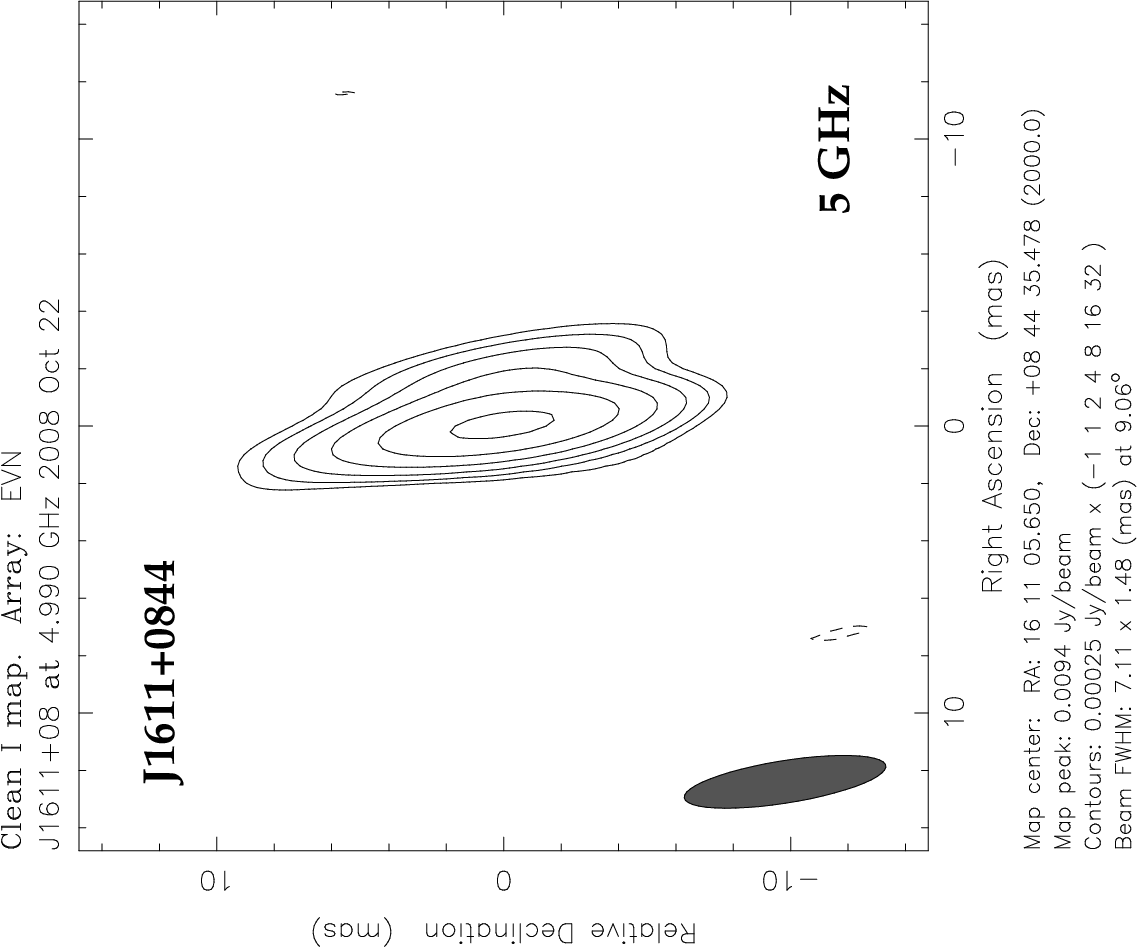}
  \includegraphics[bb=68 170 520 626,width=75mm,angle=270,clip= ]{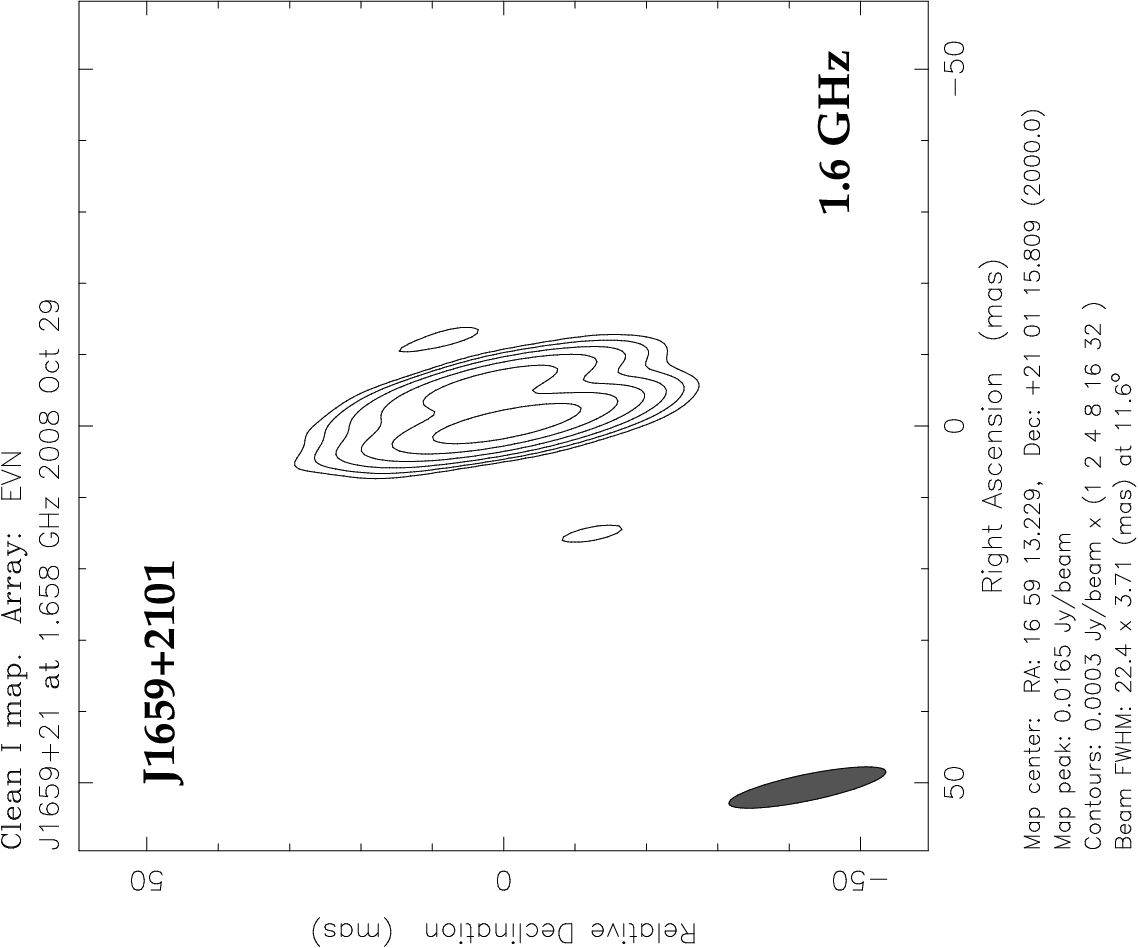}
  \includegraphics[bb=68 170 520 626,width=75mm,angle=270,clip= ]{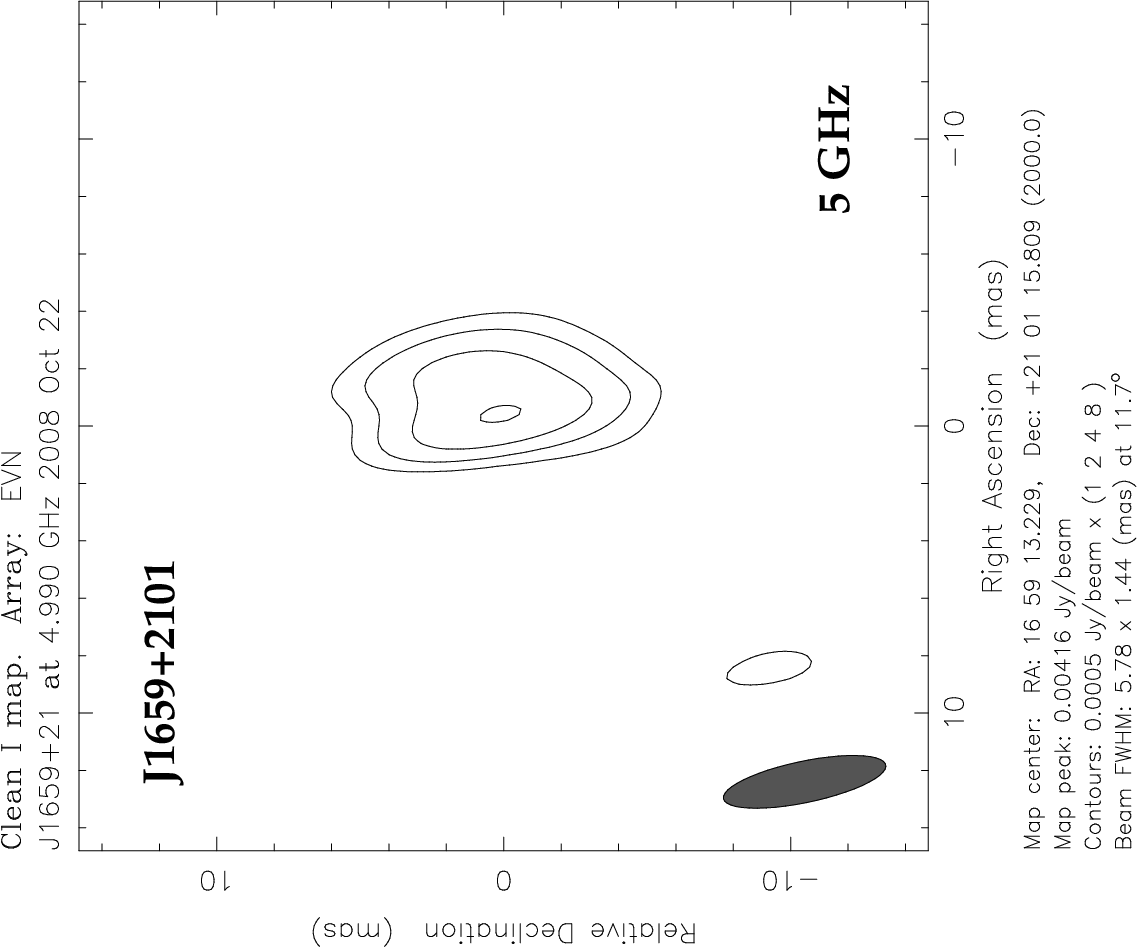}
  \caption{continued.}
\end{figure*}

\section{Results}
\label{results}

The five targeted high-redshift quasars were clearly detected and imaged with the EVN at both 1.6~GHz and 5~GHz (Fig.~\ref{images}). However, we could not detect any mas-scale compact radio source at the position of the nearby ($6\farcs8$) FIRST radio companion (Fig.~\ref{first}) to J0813+3508, at a brightness level of 0.3~mJy/beam ($3\sigma$). This indicates that the secondary source is extended and thus completely resolved in our VLBI experiments; notably, the position angle of the jet-like structure in our VLBI images of J0813+3508, especially at 1.6~GHz (Fig.~\ref{images}, top left) practically coincides with that of the FIRST companion. Based on this directional coincidence, we believe that there is a physical association between the two radio sources. This is further supported by the fact that the visibility amplitudes on the shortest and most sensitive baseline from Effelsberg to Westerbork were somewhat higher than could be fully accounted for with the CLEAN model components derived from the full VLBI array data. From poorly constrained Gaussian model fitting, there was a hint on an extended component about $0\farcs6$ away from the compact ``core'' to the northwest in the continuation of the $\sim$10-mas scale jet and in the direction of the FIRST radio companion. 

The coordinates of the brightness peaks are estimated from the 5-GHz images using the AIPS task {\sc maxfit} and listed in Table~\ref{coords}. The uncertainties are determined by the phase-reference calibrator source position accuracy, the target--calibrator angular separation, the angular resolution of the interferometer array, and the signal-to-noise ratio.

Difmap was used to fit circular Gaussian brightness distribution model components to the interferometric visibility data of the five quasars detected. We calculated the two-point spectral indices for the bright central components from the model flux densities. From the 5-GHz data, we also derived the brightness temperatures of these dominant components as
\begin{equation}
T_{\rm B}=1.22 \times 10^{12} (1+z) \frac{S}{\vartheta^{2} \nu^{2}} \,\,\,[\rm{K}],
\end{equation}
where $z$ is the redshift, $S$ the flux density (Jy), $\nu$ the observing frequency (GHz), and $\vartheta$ the full width at half maximum (FWHM) size of the Gaussian measured in mas (e.g. Condon et al. \cite{cond82}). The model parameters, and the values derived from them are given in Table~\ref{modelfit}.  

\begin{table}
  \caption[]{Measured coordinates of the five radio quasars at $z>4.5$ imaged with VLBI.}
  \label{coords}
\begin{tabular}{lcccc}        
\hline\hline                 
Source name & Right ascension   & Declination  & \multicolumn{2}{c}{Uncertainties}  \\
            & $^{\rm h}$ $^{\rm m}$ $^{\rm s}$ & $\degr$ $\arcmin$ $\arcsec$   & mas & mas \\ 
\hline                       
J0813+3508  & 08 13 33.32789  & 35 08 10.7698  & 0.4 &  0.5 \\
J1146+4037  & 11 46 57.79043  & 40 37 08.6256  & 0.3 &  0.4 \\
J1242+5422  & 12 42 30.58994  & 54 22 57.4524  & 3.9 &  1.9 \\
J1611+0844  & 16 11 05.65000  & 08 44 35.4776  & 0.5 &  0.4 \\
J1659+2101  & 16 59 13.22857  & 21 01 15.8087  & 1.0 &  1.1 \\
\hline   
\end{tabular}
\\
Notes: Col.~1 -- source name; Col.~2 -- J2000 right ascension ($^{\rm h}$ $^{\rm m}$ $^{\rm s}$); Col.~3 -- J2000 declination ($\degr$ $\arcmin$ $\arcsec$); Col.~4 \& 5  -- estimated positional uncertainties in right ascension and declination (mas).
\end{table}

\begin{table*}
  \caption[]{The parameters derived for the quasars from VLBI imaging.}
  \label{modelfit}
\begin{tabular}{ccccccccc}        
\hline\hline                 
Source name & \multicolumn{2}{c}{Complete VLBI flux density (mJy)} & \multicolumn{2}{c}{Main component flux density (mJy)}  & $\alpha$ & Size & $T_{\rm B}$ & $L$ \\
            & at 1.6~GHz    & at 5~GHz                                & at 1.6~GHz    & at 5~GHz                &          & mas       & $10^{9}$~K  & $10^{26}$~W~Hz$^{-1}$\\ 
\hline                       
J0813+3508  & 17.1$\pm$0.9 &  7.3$\pm$0.4$^{*}$ & 12.8$\pm$0.6  &  6.8$\pm$0.3$^{*}$  & $-0.6$   & 1.18$\pm$0.02  &  1.5$\pm$0.1   &  8.4$\pm$0.4  \\
J1146+4037  & 15.5$\pm$0.8 &  8.6$\pm$0.4$^{*}$ & 15.5$\pm$0.8  &  8.4$\pm$0.4$^{*}$  & $-0.5$   & 0.74$\pm$0.01  &  4.5$\pm$0.3   &  10.7$\pm$0.5 \\
J1242+5422  & 17.7$\pm$0.9 &  9.7$\pm$0.5 & 17.1$\pm$0.9  &  9.4$\pm$0.5              & $-0.5$   & 0.67$\pm$0.01  &  5.9$\pm$0.5   &  9.0$\pm$0.5  \\
J1611+0844  & 13.0$\pm$0.8 & 12.9$\pm$0.6 & 13.0$\pm$0.8  & 12.4$\pm$0.6              & $-0.0$   & 0.85$\pm$0.01  &  4.7$\pm$0.3   &  4.6$\pm$0.2  \\
J1659+2101  & 29.3$\pm$1.5 & 10.6$\pm$0.7 & 19.7$\pm$1.0  & 10.6$\pm$0.7              & $-0.6$   & 3.07$\pm$0.12  &  0.3$\pm$0.04  &  12.3$\pm$0.8 \\
\hline   
\end{tabular}
\\
Notes: 
Col.~1 -- source name; Col.~2-3 -- complete VLBI flux density at 1.6~GHz and 5~GHz (mJy); 
Col.~4-5 -- main component VLBI flux density at 1.6~GHz and 5~GHz  (mJy);
Col.~6 -- two-point spectral index of the main component;
Col.~7 -- fitted circular Gaussian diameter (FWHM) for the main component at 5~GHz (mas); 
Col.~8 -- rest-frame brightness temperature at 5~GHz ($10^{9}$~K);
Col.~9 -- rest-frame monochromatic 5-GHz luminosity ($10^{26}$~W~Hz$^{-1}$). 
The VLBI flux density calibration uncertainties are assumed as 5\%. The statistical errors of the fitted model parameters are estimated according to Fomalont (\cite{Foma99}).\\
$^{*}$ flux density value corrected for an estimated coherence loss of 5\%
\end{table*}

\section{Discussion} 
\label{discussion}

All of the $z>4.5$ quasars we imaged are somewhat resolved, often with structures extending up to several 10~mas (Fig.~\ref{images}). For comparison, a 1~mas angular size corresponds to 6.3$-$6.6~pc projected linear size, depending on the actual redshift of the object. The measured brightness temperatures ($T_{\rm B}$$\sim$$10^8-10^9$~K; Table~\ref{modelfit}) clearly indicate AGN activity, but are at least an order of magnitude lower than the equipartition value estimated for relativistic compact jets ($T_{\rm B,eq} \simeq 5 \times 10^{10}$~K; Readhead \cite{read94}). It may suggest that $(i)$ in the radio spectrum of the sources we probe regions far away from the peak frequency caused by synchrotron self-absorption; $(ii)$ the jet viewing angles are
at least moderate, and in fact we experience Doppler-deboosting of the emission; or $(iii)$ the intrinsic brightness temperatures of compact jets at very high redshift are significantly lower than the equipartition value.
Except for one flat-spectrum source (J1611+0844), the spectra of the quasar ``cores'' (i.e. the main components) in our sample are steep ($\alpha \leq -0.5$; Table~\ref{modelfit}). The intrinsic brightness temperatue was determined for the blazar J1430+4204 ($z=4.72$) by comparing the variability brightness temperature and the one measured with VLBI. It was found that $T_{\rm B,int} \simeq T_{\rm B,eq}$ (Veres et al. \cite{vere10}), so there is no reason to assume that the intrinsic brightness temperatures are generally lower at $z>4.5$. 

A possible signature of the blazar nature of a source is its strong variability. We can look for an indication of long-term variability, if we compare the 1.4-GHz flux densities taken from the VLA FIRST catalogue (Table~\ref{targets}, last column) with our 1.6-GHz flux densities measured in our EVN run (Table~\ref{modelfit}). Since our angular resolution is much higher, if a source is stationary, then one would expect the VLBI flux densities to be lower than or at best equal to the VLA values within the uncertainties. This is the case for all our sources but J1611+0844.
That J1611+0844 -- the only flat-spectrum source in our sample -- has nearly 50\% higher flux density measured at a later epoch at much higher resolution proves its non-stationarity. 
Both the variability and the flat spectrum argue for Doppler boosting, but the measured brightness temperature of this source does not support it. In the case of J1611+0844, the observed one-sided compact core--jet structure at 5 GHz and the quasar classification can just be made consistent with the simple orientation-dependent unified picture of radio-loud AGNs (e.g. Urry \& Padovani \cite{Urry95}). The apparent deboosting (Doppler factor $\delta=T_{\rm B}/T_{\rm B,eq}\simeq0.1$) can be explained if, e.g., we assume a $\theta=45\degr$ jet angle to the line of sight and an extremely (but not impossibly) large bulk Lorentz factor ($\gamma$$\simeq$$35$).

The case of J0813+3508 is similarly puzzling. If, as our data suggest, its FIRST companion (Fig.~\ref{first}) is associated with the quasar at $z=4.92$, the projected linear size of the source is 43~kpc. This means that J0813+3508 could be the quasar with the most extended radio jet known at an extremely high redshift (cf. Cheung et al. \cite{cheu05,cheu08}). The markedly one-sided arcsecond-scale radio structure suggests Doppler boosting and a small jet angle to the line of sight. However, the measured $\sim$$10^9$~K brightness temperature and the steep spectrum of the compact VLBI component do not support this view. It is possible that there is a significant misalignment between the mas-scale and arcsecond-scale jet, or -- less probably -- the jet in this quasar is intrinsically one-sided. An alternative scenario is that we see an expanding and extended radio source with double hot spots, of which the nearest and approaching one is detected with VLBI. Future sensitive radio interferometric observations at intermediate angular resolutions could help reveal the true nature of this object (or these objects).     
 
The remaining three sources (J1146+4037, J1242+5422 and J1659+2101) are quite alike in their compact structure (Fig.~\ref{images}), steep spectra, and luminosity (Table~\ref{modelfit}). The quasar J1659+2101 is somewhat more resolved than the others. Notably, the two most distant ($z$$\sim$6) quasars (J0836+0054, Frey et al. \cite{frey03,frey05}; J1427+3312, Frey et al. \cite{frey08b}, Momjian et al. \cite{momj08}) observed with VLBI to date share much similar properties.

The rest-frame 5-GHz luminosities of the main components of our sources ($L$$\sim$$10^{26}-10^{27}$~W~Hz$^{-1}$; Table~\ref{modelfit}) are comparable to the values of a large sample of gigahertz peaked-spectrum (GPS) and compact steep-spectrum (CSS) sources compiled by O'Dea (\cite{Odea98}). Our values refer to the VLBI components and can therefore be regarded as lower limits to the radio luminosities of the whole sources. We found no evidence of any significantly Doppler-boosted radio emission in our cases, so that the measured luminosities indicate intrinsically powerful sources. 

GPS and CSS sources are thought to be young, evolving objects, and/or perhaps ``frustrated'' ones that are confined by the dense ambient gas (see O'Dea \cite{Odea98} for a review). A subclass of these sources are the CSOs, which show nearly symmetric double or triple morphology when imaged at VLBI resoulution. According to the model of Falcke et al. (\cite{Falc04}), our high-redshift steep-spectrum objects may represent GPS sources at early cosmological epochs. The first generation of supermassive black holes could have powerful jets that developed hot spots well inside their forming host galaxy, on linear scales of 0.1$-$10~kpc. Taking the relation between the source size and the turnover frequency observed for GPS sources into account, and for hypothetical sources matching the luminosity and spectral index of ours, Falcke et al. (\cite{Falc04}) predict that the angular size of the smallest ($\sim$$100$~pc) of these early radio-jet objects would be in the order of 10~mas at $z$$\sim$$5$, and the observed turnover frequency in their radio spectra would be around 500~MHz. These angular sizes are indeed seen in our VLBI images (Fig.~\ref{images}). To confirm the low-frequency turnover, high-resolution interferometric flux density measurements would be needed at multiple frequencies down to the 100-MHz range -- a task well-suited to the Square Kilometre Array (SKA) now under development, in its high-resolution configuration. In a sense, the observed spectral properties of high-redshift sources are a resurrection of the indication on ``humped'' spectra of high-redshift quasars found more than 20 years ago when the highest known redshift was below 4 (Peterson et al. \cite{pete82}; O'Dea \cite{odea90}).

Currently we do not see many blazars at $z$$\sim$5 or higher. According to Table~\ref{z4.5-sources}, the most distant blazar imaged with VLBI to date (J0906+6930; Romani et al. \cite{roma04}) has a redshift of 5.47. On the other hand, one would naively expect highly relativistically beamed, flat-spectrum radio sources to dominate the high-redshift VLBI samples. Whether the apparent absence of very high-redshift blazars is a result of some selection effect (i.e. the lack of redshift measurements), the poor statistics due to the small sample, or has a real physical cause, is to be addressed with further observations. Indeed, there are indications that blazars are not very common at the highest redshifts. Based on the evolving gamma-ray luminosity function, Inoue et al. (\cite{inou10}) expect that the Fermi Gamma-ray Space Telescope will find a few (i.e. the order of unity) blazars at $z$$\sim$6 over a period of 5 years. For the strong sources, there is a positive correlation between the gamma-ray and the parsec-scale synchrotron radio emission, with the gamma-detected sources having on average higher brightness tempeartures in the radio (e.g. Kovalev et al. \cite{kova09}). Thus the Fermi Large Area Telescope (LAT) potentially selects the brightest radio sources that would be natural targets for follow-up with VLBI.

\section{Conclusions}
\label{conclusion}

We imaged five distant radio quasars at $4.5 < z < 5$ with the EVN at two frequencies (1.6 and 5~GHz), almost doubling the currently known sample of quasars imaged with VLBI at $z>4.5$. The phase-referenced observations allowed us to derive accurate astrometric positions for our targets. The slightly resolved mas- and 10-mas-scale radio structures, the measured moderate brightness temperatures ($\sim$10$^8$$-$10$^9$~K), and the steep spectra in all but one case suggest that our sample of compact radio sources at $z>4.5$ is dominated by objects that do not resemble blazars. One of the quasars (J0813+3508) is likely to be extended to $\sim$7$\arcsec$ which corresponds to a 43~kpc projected linear size. 
It is possible that we see young, evolving, and compact GPS-like objects that are signatures of early galaxy formation where the expanding powerful synchrotron radio sources are confined by the dense interstellar medium.  

\begin{acknowledgements}
The EVN is a joint facility of European, Chinese, South African, and other radio astronomy institutes funded by their national research councils. 
This work has benefitted from research funding from the European Community's Sixth Framework Programme under RadioNet R113CT 2003 5058187, the Hungarian Scientific Research Fund (OTKA, grant no.\ K72515), and the European Community's Seventh Framework Programme under grant agreement no. ITN 215212 ``Black Hole Universe''. 
\end{acknowledgements}

\end{document}